\documentclass[prc,aps,showpacs,floatfix,nofootinbib,twocolumn,letterpaper]{revtex4} 
\usepackage{graphicx} 
\usepackage{amssymb} 
\usepackage{amsmath} 
\usepackage{amsfonts}

\def\be{\begin{equation}} 
\def\ee{\end{equation}}

\def\gb{{ \tilde \gamma}}
\def\ve{\varepsilon}
\begin{document} 
\title{A schematic reaction-theory model for nuclear fission
} 
\author{G. F. Bertsch } 
\affiliation{
Department of Physics and Institute of Nuclear Theory, 
Box 351560\\ University of Washington, Seattle, Washington 98915, USA} 
\email{bertsch@uw.edu} 
\begin{abstract} 
 
The K-matrix formalism is applied to a schematic model for nuclear fission.
The purpose is to explore the dependence of observables on the assumptions
made about the configuration space and nucleon interaction in the
Hamiltonian of the fissile nucleus. As expected,
branching  ratios in induced fission are found to depend sensitively on 
the character of the residual interaction, whether it is pairing in form
or taken from a random ensemble. On the
other hand, the branching ratio is not much affected by the presence of
additional configurations that do not introduce new fission paths.

\end{abstract}

\maketitle 
 
\section{Motivation}

Nuclear fission is one of the most challenging topics in the
quantum theory of finite many-particle systems.  Useful
phenomenological models are available that take into account
both nuclide-dependent shell effects and nucleon-blind
collective variables, see \cite{sch16,sch18} for recent reviews. 
But anchoring these models to the underlying many-body
Hamiltonian faces enormous obstacles related to the huge number
of many-body configurations participating in the dynamics.

It is obvious that the practical theory should be informed by
microscopic Hamiltonian dynamics, but since a full Hamiltonian
theory is presently out of reach, it might be useful to examine
simple models that include qualitative aspects of the complete
Hamiltonian.  Perhaps the reliability of the various
approximation schemes can be assessed in much smaller spaces than
would be required for a quantitative theory.  It is the goal of
the present work to propose a simplified model for this purpose.

Most fission theory based on nucleonic Hamiltonians is carried
out in the time domain, for example with time-dependent 
mean field approximations\cite{sim18,bul15} .  In contrast, the physical
observables in fission reactions are the energy-dependent cross sections.
This is another reason for using reaction theory in  constructing models.

\section{Reaction theory}
\label{model}
The K-matrix formalism is well-suited for a Hamiltonian-based reaction
theory of multi-particle systems\footnote{This is in contrast to the 
$R$-matrix theory \cite{wig47} which is convenient for phenomenological models
but is not easy to apply at the level of realistic Hamiltonians. See Ref.
\cite{bou13} for a recent application to fission.}.  It has been used in a
number of different branches of physics\cite{dal61,chu95,fyo96,alh00,lin19}.  
In nuclear physics,
it has been applied nucleon-induced reactions using
Hamiltonians based on nucleon-nucleon interaction\cite{mah69}.  It has also
been successfully applied to develop statistical reaction
theory\cite{mit10,kaw15}.  The $K$-matrix
formalism is built on two matrix components.  The first is a Hamiltonian
matrix $H$ in the space of internal or quasi-bound
configurations. Some of
these configurations have decay amplitudes to possible
final-state channels. These amplitudes
are contained in a second matrix $\gb$.  It has $N_{conf}$ rows
corresponding to the dimension of $H$ and $N_{ch}$ columns
corresponding to the number of decay channels.  The $K$-matrix
is defined as

\be
K =  \pi{ \gb}^T \frac{1}{ E - H} { \gb}
\ee 
where $E$ is the total energy of the reacting system.  The
$S$-matrix is computed as
\be
S = \frac{1 -i  K}{  1 +i  K}.
\label{S} 
\ee
The partial width $\gamma_{\mu c}$ for a configuration $\mu$ decaying through 
channel $c$  is
\be
\Gamma_{\mu c} = 2 \pi \gamma^2_{\mu c}.
\ee

Eq.~(1) effectively separates the computational problem into
separate tasks. The first task is the construction of a Hamiltonian
matrix\footnote{More rigorously, the matrix $H$ includes
the level shifts due to coupling to continuum channels. In
practice, these shifts are small and can be ignored.
It should also be mentioned that Eq. (\ref{S}) as given neglects
effects of the scattering potentials on the elastic phase
shifts within the individual channels.  
They are straightforward to include but are not needed for inclusive cross
sections.} $H$ for the internal states.  It requires setting up 
a basis composed of many-body configurations and computing
the interactions between the configurations.  
This configuration-interaction (CI) approach is very well known, and
it has been very successful in many fields including nuclear structure 
physics.  The second
task is to calculate $\gb$,
the matrix of partial
decay widths of the internal states to the continuum channels.
This is much more challenging when the channel states are all
composite particles; at this point the needed approximations are 
not testable with simple models \cite{ber19}.  Given the two matrices, all
that remains of the computation is ordinary linear algebra.

\section{Model Hamiltonian}

The requirements for the model Hamiltonian are that it:\\
--is expressible in terms of one- and two-body Fock-space operators;\\
--defines an operator $\hat Q$ that can be used to measure the
evolving shape of the fissioning system;\\
--is flexible enough to simulate induced fission as well as spontaneous 
fission.

These requirements can be fulfilled by the following model.
Configurations are generated in a 
space of $N_{orb}$ orbitals, each orbital containing two time-reversed pairs
$k$ and $\bar k$.
The first $N_{orb}/2$ orbitals are fully occupied in the
ground-state configuration, while the remaining $N_{orb}/2$
orbitals are fully occupied in the doorway state to
fission.  
In operator representation, the Hamiltonian is 
\be
\hat H = \sum_{k=1}^{N_{orb}/2} \left((k\; \mathrm{mod}\;
N_{orb}/2 )-1\right) \varepsilon_0 \hat
n_{k}  
+ v_Q \hat Q \hat Q 
+ \sum_{k,k'} v_{k,k'} \hat P^\dagger_{k}\hat P_{k'}.
\label{ham-eq}
\ee
Here
$\hat n_{k} = \hat a^\dagger_{k}  \hat a_{k}
+\hat a^\dagger_{\bar k} \hat a_{\bar k}$ 
gives the occupation number of the $k$ orbital,
$\hat P^\dagger_{k} = \hat a^\dagger_{k} \hat a^\dagger_{\bar k}$ is the 
pair creation operator, and $\hat Q = \sum_{k} q \hat n_{k}$ 
measures the elongation of the configuration. In the expression for $\hat Q$
$q(k) = \pm 1$ for the first $N_{orb/2}/2$ orbitals and $-1$ for the
others.  Note that the first two terms in $\hat H$ are diagonal in the
configurations and serve to define their energies. 

The qualitative scheme of orbital energies 
and how they are
filled for low-energy configurations is shown in Fig. 1, assuming
the parameter $v_Q$ in the Hamiltonian is negative.
\begin{figure}[tb] 
\begin{center} 
\includegraphics[width=\columnwidth]{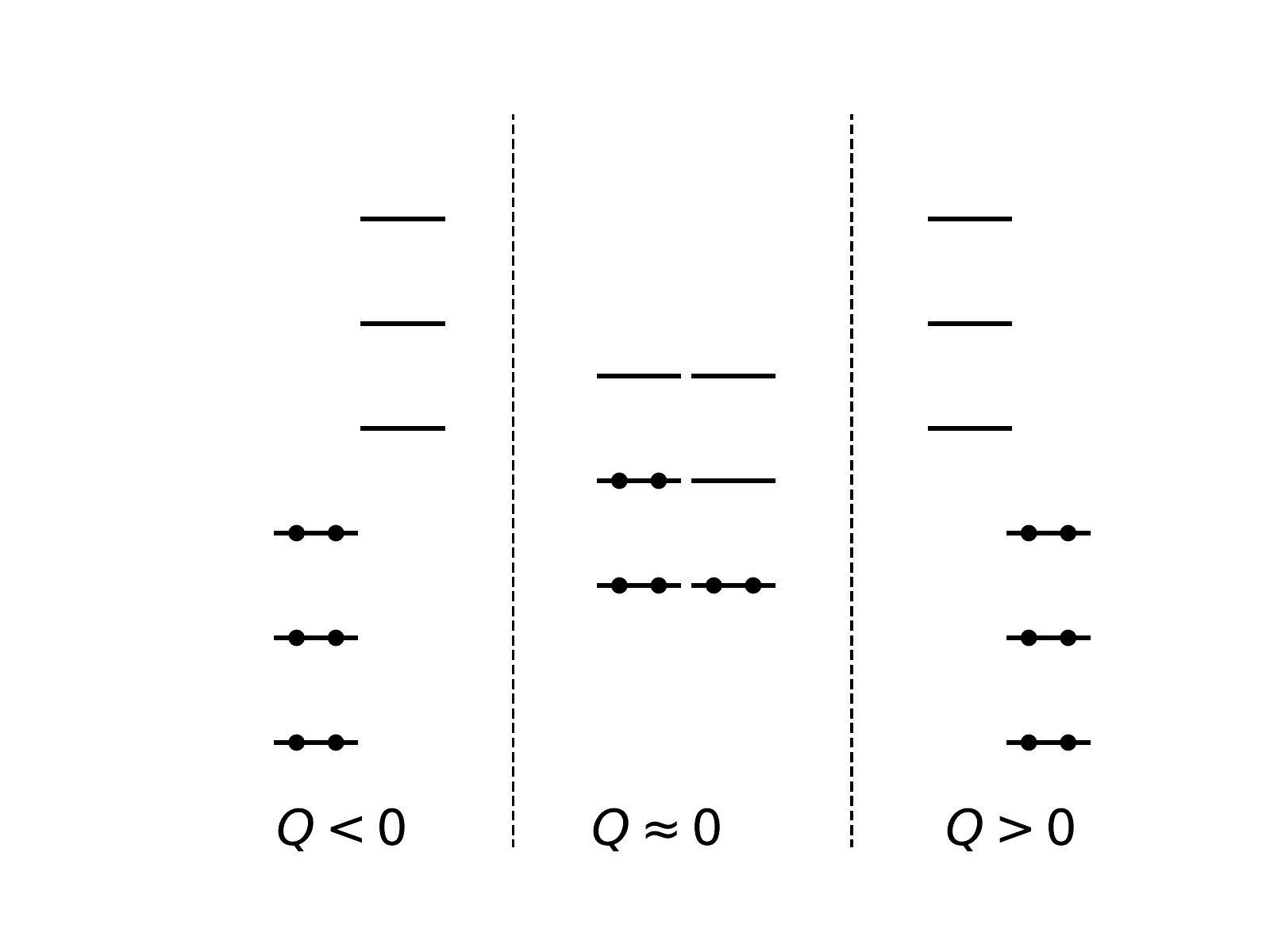} 
\caption{Single-particle spectra 
at different deformations, taking $v_Q<0$ in the model
Hamiltonian.
The orbital
occupancies of low-energy configurations are indicated by the
filled circles.
}
\end{center}
\end{figure}
The left-hand configuration has all $q=-1$ orbitals filled 
and represents the ground state of 
the fissile nucleus.  The one on the right with all $q=+1$
orbitals filled represents a scission doorway configuration.

For the numerical calculations we take $N_{orb} = 6$ for the dimension
of the orbital space.  The space is half filled with $N=6$ nucleons that 
occupy the orbitals as pairs.  The configurations are thus restricted
to seniority zero; the dimension of the space is
\be
N_{conf} = { 6 \choose 3} = 20.
\ee 
$Q$ in the model ranges 
from -6 to +6, with one state at each end point and 18 states 
in between.
As an example, Fig. 2 
\begin{figure}[tb] 
\begin{center} 
\includegraphics[width=1.0\columnwidth]{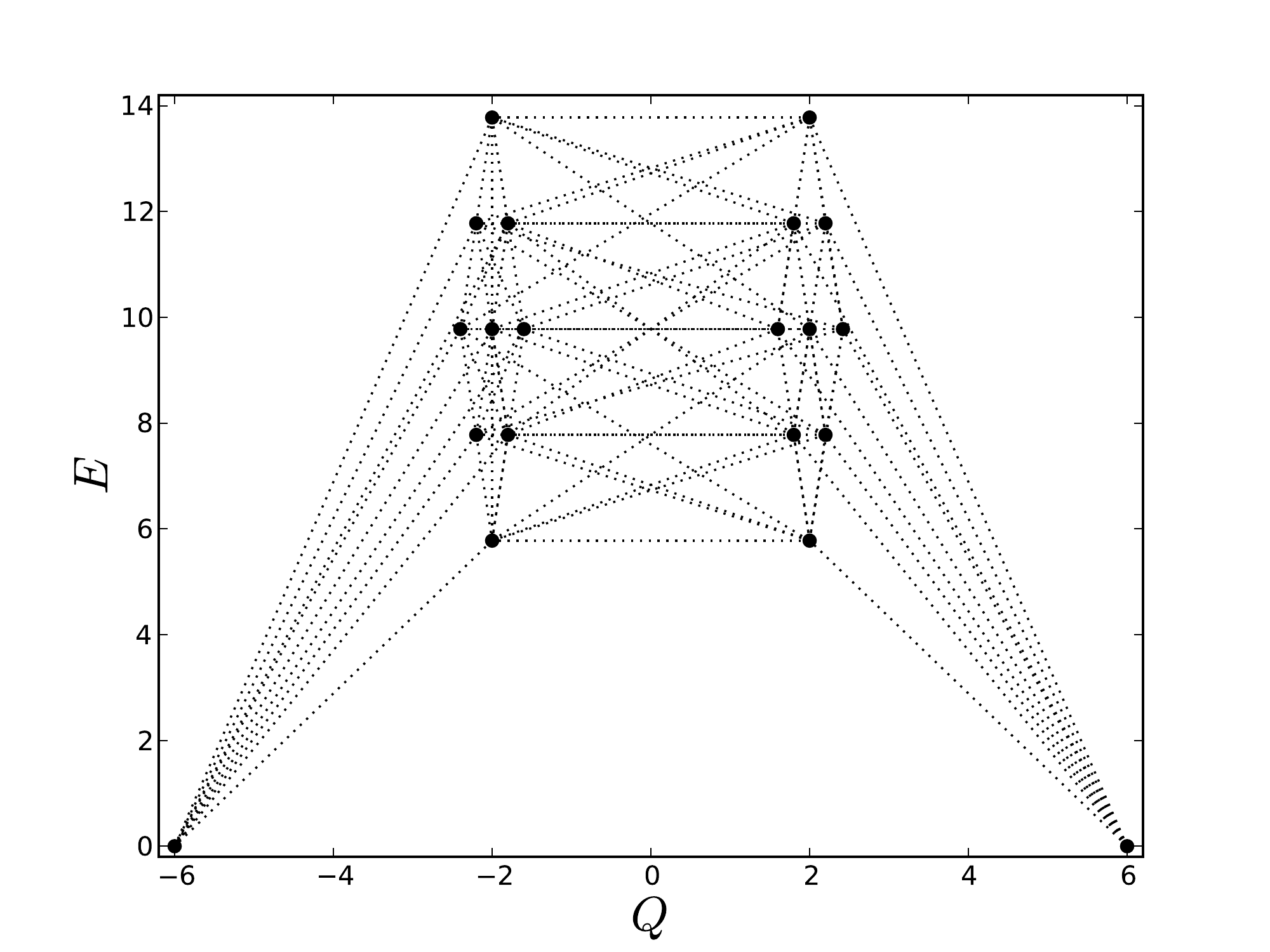}
\caption{
Spectrum of the barrier model.
The vertical axis has been shifted to
put the entry and prescission configurations at zero energy.
The horizontal
positions of the configurations have been slightly shifted to 
make visible the degeneracies of the intermediate configurations.
The dotted lines 
join configurations that are connected by $v_{k,k'}$.  
}
\label{barrier-fig}    
\end{center} 
\end{figure} 
shows the energy  of the configurations as a function of
$Q$ in the "barrier" model, constructed with $v_Q = -\frac{1}{4} e_0$. The Figure
also shows the connectivity of the network linked by the two-particle
interaction.  The state at  
$Q=-6$ will be coupled to an entrance channel, and the one at $Q=+6$
will be coupled to a fission channel.
With the energies of the configurations as depicted in Fig. 2,
the model could simulate the fission
cross section in the presence of a barrier along the fission
path.  Note that $Q$ is
a discrete property of individual configurations.  This is to be
contrasted with the generator-coordinate method (GCM) of
constrained Hartree-Fock theory\cite{ben03} which treats the expectation value
$\langle Q\rangle$ as a continuous variable.

For the present study we will examine the fission-to-capture branching 
ratio at energies above the barriers.  Physically, the level density of internal
states is very large.  In the model with $V_Q=0$, the highest level density of the
internal states is at the entry doorway energy.
The resulting configuration energies are shown in Fig. 3; we
shall call this the ``no-barrier" model.  Note that the fission
doorway energy is also at the point of highest level density.
\begin{figure}[tb] 
\begin{center} 
\includegraphics[width=\columnwidth]{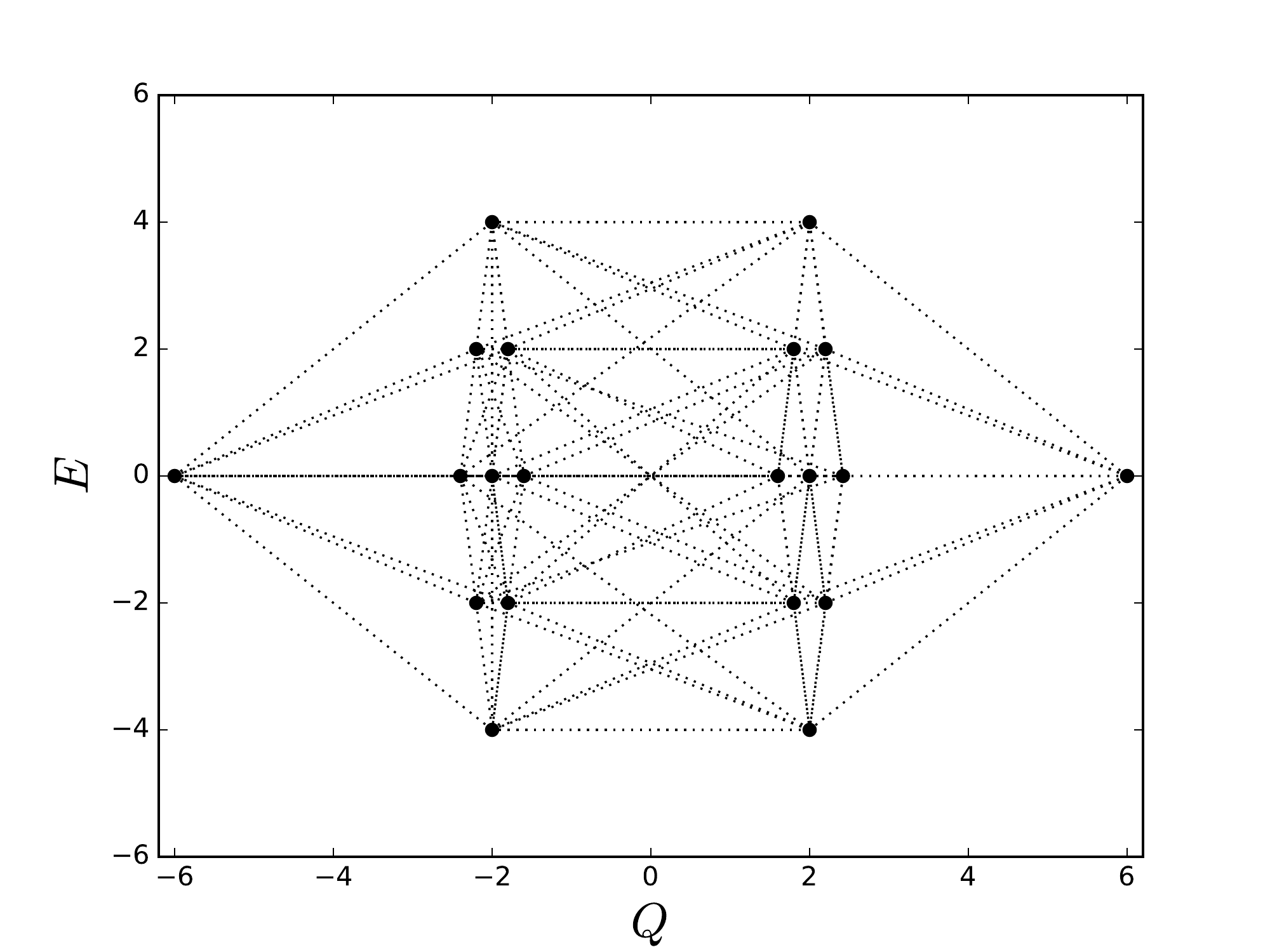} 
\caption{Spectrum of the "no barrier"  Hamiltonian.
The vertical axis has shifted to
put the first and last configurations at zero energy.  
}
\label{induced-fig}
\end{center}
\end{figure}

Two extreme choices for the interaction
will be examined.  The first is the pure pairing model,
\be
v_{k,k'} = -G  \,\,\,\,\,\, {\rm (pairing)}
\label{pairing}
\ee
where $G$ is the pairing strength.
The other extreme is a random interaction taken from a Gaussian ensemble.
Here the probability $W$
of the interaction strength  $v_{k,k'}$ is taken as
\be
\label{random}
W(v_{k,k'}) = \frac{1}{\sqrt{2 \pi G^2}}\exp(-v_{k,k'}^2/2G^2)
\,\,\,\,\,\,{\rm (random)}.
\ee
The two model interactions have equal  rms  matrix elements.

There is a technical problem in using the Hamiltonian as given
for the pairing interaction.  Namely, the uniform spacing of the
single-particle energies produces a significant
degeneracy in the energies of the configurations.  This might
give rise to unphysical effects in the transport 
properties of the Hamiltonian. This problem is mitigated in the
numerical calculations by modifying the single-particle energies
to
\be
\ve_k' = \ve_k  + 0.1 r_k e_0.
\label{vek}
\ee
Here $r_k$ is a random number of unit variance taken from a 
Gaussian ensemble.  In fact this complication of the model is
physically warranted: the spacings are also not uniform 
in more realistic models. 
The cost of introducing random terms into the Hamiltonian is that
the ensemble must be sampled multiple times to compute
observables.

The two-particle interaction strength is the last parameter in
the Hamiltonian that needs to be set.
Here one can get guidance from empirical pairing systematics.
The mixing between low-lying Hartree-Fock
configurations is controlled by the ratio of pairing gap $\Delta$ 
and the average level spacing $e_0$ of the single-particle orbitals.
In actinide  nuclei, the (neutron) pairing gap is about
\be
\Delta \approx 12./A^{1/2} \approx  0.75 \;\;\mathrm{MeV}.
\ee   
The neutron
orbital spacing roughly given in terms of the number of neutrons $N$ in the
nucleus and their kinetic energy at the Fermi surface $\varepsilon_f$
as 
\be
e_0 \approx \frac{4 \varepsilon_f}{3 N} \approx 0.33  \;\; \mathrm{MeV}.
\ee 
This yields a ratio $\Delta/\ve_0 \approx 2.25$.    This is close to the calculated ratio for the barrier model 
taking the pairing strength to be $G=e_0$;  this value is adopted for numerical
computation in the following section. 
Table \ref{params} summarizes the numerical parameters for the no-barrier
model.
\begin{table}[htb]
\begin{center}
\begin{tabular}{|c|c|}
\hline
Parameter &  Value \\
\hline
$N_{orb}$ & 6\\
$N$  & 6\\
$v_Q $ & 0  \\
$ G $ & 1  \\
\hline
\end{tabular}
\caption{Parameters for the model Hamiltonian.
Energies are
in units of $e_0$.}
\label{params}
\end{center}  
\end{table}

\section{Application to branching ratios}

In this section the model is applied to the fission-to-capture branching 
ratio. To treat this as a reaction in the K-matrix theory,
we need to assign partial widths for the entrance, capture, and
fission channels.  For a fissile nucleus such
as $^{235}$U bombarded with low-energy neutrons, the fission 
and capture widths are comparable, and the entrance channel 
width is small compared to both of them.  This is achieved
by the partial widths assignments shown in Table II.
\begin{table}[htb]
\begin{center}
\begin{tabular}{|c|c|c|}
\hline
Configuration $\mu$ & Channel  &  $\Gamma_{\mu,c}/e_0$ \\
\hline
$ 1 $ & n  & $0.063$  \\
$1$ &  c & $1.0$  \\
$ 20$ & f & $1.0$ \\
 \hline
\end{tabular}
\caption{Partial widths for modeling the fission-to-capture branching ratio.
Channel labels are: $n$ for entry channel as in neutron-induced fission;
$c$ capture leading to the ground state of the fissile nucleus; $f$
fission decay.
}
\label{widths}
\end{center}  
\end{table}    

We can now apply the $K$-matrix formula to calculate the $S$-matrix
elements for the three channels.  The cross sections show 
large fluctuations associated with individual resonances in
the internal structure of the fissioning nucleus.  This may be
seen in Fig. 3 and 4, plotting the strengths $|S_{nc}|^2$ 
and $|S_{nf}|^2$ as a function
of energy.
\begin{figure}[tb] 
\begin{center} 
\includegraphics[width=0.5\columnwidth]{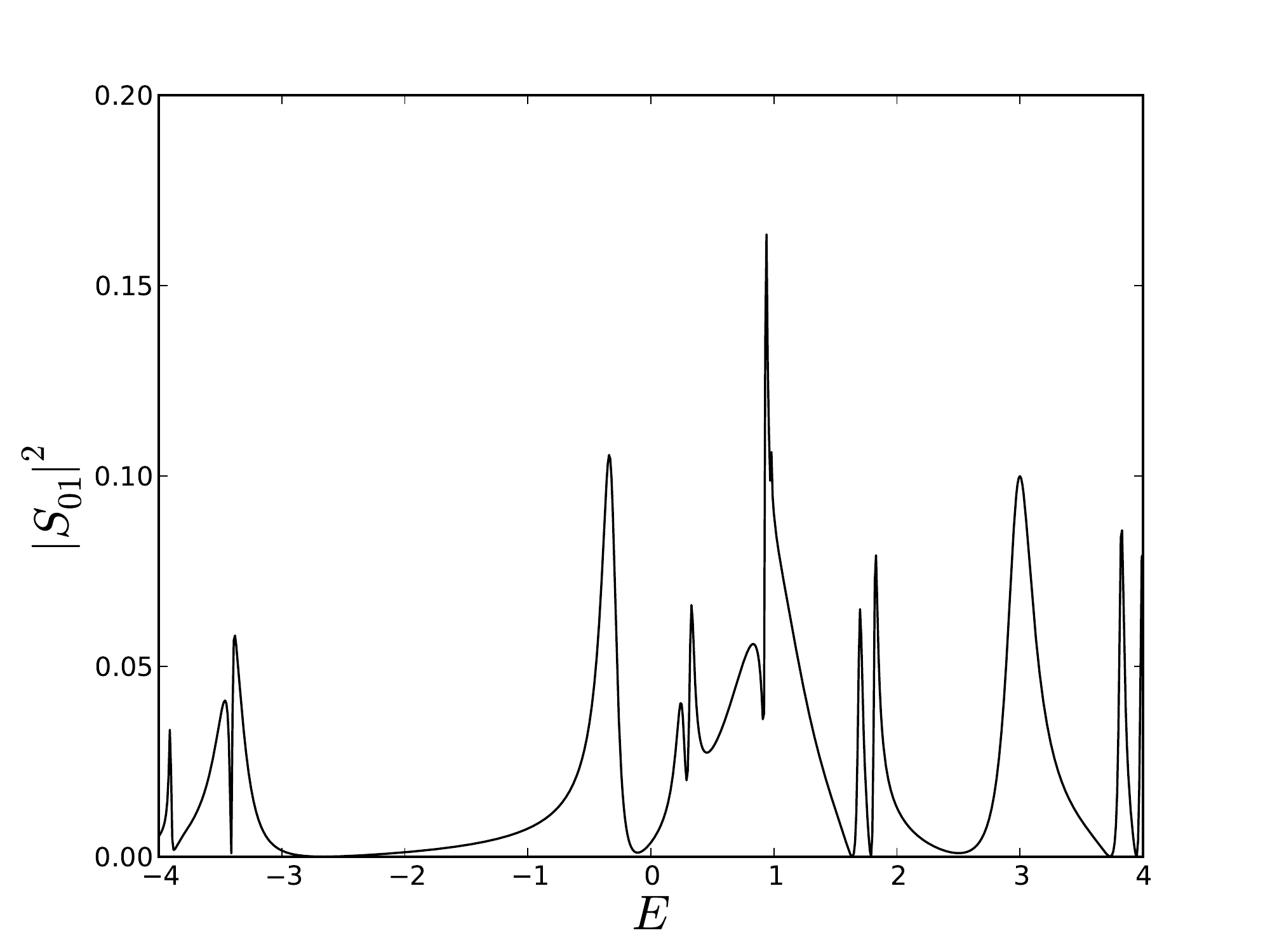}
\includegraphics[width=0.5\columnwidth]{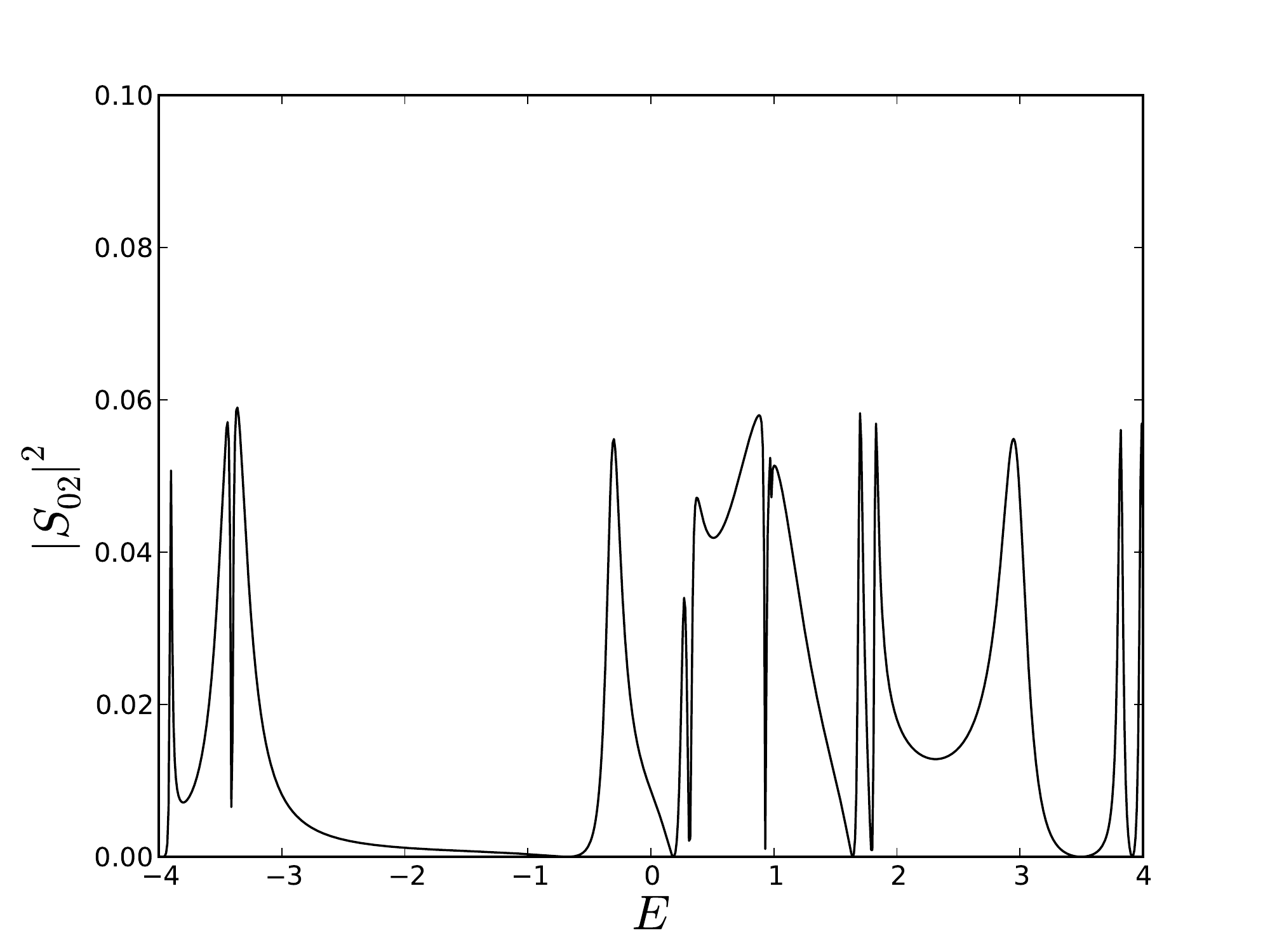}
 \caption{$S$-matrix transmission strengths for the pairing Hamiltonian.
 Upper panel:  $|S_{01}|^2$; lower panel: $|S_{02}|^2$.
}
\label{S^2}    
\end{center} 
\end{figure} 

The branching ratios are highly fluctuating quantities as a function of
energy, so we only report averages.  The fission-to-capture ratio is
calculated as
\be
\alpha^{-1} = \frac{\langle |S_{nf}|^2\rangle}{\langle |S_{nc}|^2\rangle},
\ee
where the brackets denote an energy average.
We have taken a window of energies from $E = - 4\ve_0$
to $E = 4\ve_0$ for making the averages.  The results are
shown in Table III under the column $\alpha^{-1}$.  
\begin{table}[htb]
\begin{center}
\begin{tabular}{|c|c|c|}
\hline
$N_{conf}$ &  $v_{k,k'}$  & $\alpha^{-1}$  \\
\hline
20  & pairing   & $0.82\pm0.06$ \\
20    & random  & $0.28\pm0.13$\\
20+18 & pairing  & $0.73\pm0.05$\\
20+18  & random  & $0.28\pm0.12$\\
\hline
\end{tabular}
\caption{Branching ratio $\alpha^{-1}$ for the parameter treatments
discussed in the text.  
}
\label{branching}
\end{center}  
\end{table}    
The first entry in the Table uses the pairing Hamiltonian
following Eq.~(4), (\ref{pairing}) and (\ref{random}).  The branching between capture and fission is
close to one for the chosen parameters.
This is just what one expects in the na\"ive
compound nucleus model, since the Hamiltonian has equal couplings to 
neutron capture and fission\footnote{This ignores the usual width-fluctuation
correction.}.  However, from the perspective of transport theory one would
have expected the branching to the exit channel from the entry doorway 
would be much favored. That turns out to be the case when the random interaction is used in the
model, as may be seen in the second line of Table III.
There the calculated branching ratio
 is a factor three smaller than 
the pairing or compound nucleus models.
Clearly, the coherence of the pairing interaction has a major effect on
branching ratios between different decay modes.  While that qualitative
conclusion does not come as a surprise,
the model shows that the means to study
such issues are at hand, given an adequate basis of 
configurations and their couplings to decay channels.

We would also like to see the effects of the severe model space truncation.
It is easy to add configurations that couple to the ones in the Hamiltonian
but do not change the deformation $Q$; these may be called ``spectator"
configurations. The bottom two lines in Table III
show the results for augmenting the space by adding 18 spectator states,
each one coupled to a non-doorway state of the Hamiltonian depicted in
Fig. 3.  As may be seen in Table III, the resulting branching ratio  for both interactions is hardly
changed or not changed at all.
This is good news for justifying the drastic truncations of the
configuration spaces needed in more realistic models.  However, it is somewhat puzzling that models
that couple collective variables to internal degrees of freedom,
eg. Ref. \cite{cal83,sca15},  show significant effects in time-dependent
dynamics.  It may be that branching ratios are nevertheless
insensitive to spectator configurations, or it might be that the
present size of the model space is too small to see a real effect.

\section{Extensions of the model}

To make firm conclusions about the approximations 
invoked in realistic fission theory, it is essential to
include both neutron and protons in the Hamiltonian. To
this end, one can easily generalize Eq. (\ref{ham-eq}) to include
both species of particles, allowing them to interact
with each other through the field $\hat Q$.  The dimension
of the space as constructed in Sect. (\ref{model})  increases from
$N_{conf} = 20$ to $20^2 = 400$.  The resulting model has
a small enough dimension to permit easy coding and quick
execution times on laptop computers.  One should not expect
qualitative differences in the comparisons that were
presented with the present model.  The changes in the proton shape
distributions will track closely with the neutrons, and the
separate coherences of the two pairing fields will preserve
and perhaps amplify the stronger transport through shape
changes.

More challenging for a more realistic model is to extend the
space beyond seniority zero to access quasiparticle excitations.
The statistical properties 
such as level densities depend crucially on these excitations.
In models of fission such as the Langevin dynamics, the
quasiparticles provide a thermal reservoir for energy exchanges
with the collective degrees of freedom.  In the context of 
the present model, the dimension of the space (for
one species) goes from 20 to
\be
{ 12 \choose 6} = 924.
\ee 
Including both protons and neutrons, the total dimension
becomes $\approx 10^6$.
The resulting computational problem is then well beyond the
capabilities of general-purpose linear algebra libraries
and laptop computers.  Aside from the numerical challenge,
it is far from clear how to parameterize the interaction
Hamiltonian between quasiparticles.  It is easy 
to model the pairing interaction and the $Q$-dependent
mean-field interaction, but interactions that change the
number of quasiparticles or scatter them from one set
of orbitals to another are not well understood.

A nice feature of CI models of induced fission is that the same Hamiltonian
can also be applied to spontaneous fission.  The physical observable for 
spontaneous fission is the lifetime 
or the decay rate.  
To calculate the decay widths, one simply adds partial fission widths
$i\Gamma_{mu f}/2$ to the diagonal energies of the fission doorway state
and diagonalizes
the resulting non-Hermitian Hamiltonian.
The mean lifetime of the ground state is then given by 
$\tau = \hbar/2 {\rm Im} E_{gs}$.  The tunneling physics is be
simulated by adjusting $v_Q$ to make a barrier between the leftmost
configuration and the fission doorway on the right, as shown in
Fig. 2.

For completeness, it should be mentioned that an important quantity for reaction theory in a statistical
regime is the channel transmission coefficient $T_c$.  This may be defined
empirically as $T_c = 1 - R_c$ where $R_c$ is the average reflection probability
for an incoming flux in the channel $c$.  The averaging makes sense only
if there are many doorways to the channel; if that is the case and
the average partial decay width is small, the transmission factor
can be calculates as
\be
T_c = 2\pi \frac{ \langle \Gamma_{\mu c}\rangle} {D}
\ee 
where $D$ is the average level spacing the doorways.  Obviously,
transmission coefficients are beyond the scope of the present model
since it has only one doorway for each decay model.  Whether an
extension of the model to many doorways can be achieved with parameters
justified by a nucleonic Hamiltonian remains to be seen.

\section{Summary}

In a general sense, the subject of this work was a simplified model of 
large-amplitude shape changes in a fermionic system.
The model can only be solved numerically,
but the dimension is small enough to carry out with desktop
tools.  The first finding is confirmation of the accepted
wisdom that the pairing interaction
plays a major role in nuclear fission, although it was not so evident in
previous models of induced fission.   With enough excitation energy,
the coherence of the pairing interaction should disappear and the
observables should be close to those calculated with the random interaction.
It would be of interest see 
what the energy limits are and how they correlate with the
collapse of the pairing condensate at finite temperature.

Another provocative finding concerns the role of spectator
configurations. The  na\"ive expectation is not borne out that these configurations
would decrease the branching ratio of fission to capture
because they would slow down the dynamic evolution.  According
to the model, that effect is quite small.  Whether it remains
small in larger and more realistic model spaces is another interesting open
question.

\section{Acknowledgments} 
I would like to thank J.~Dobaczewski, K.~Hagino, W. Nazarewicz,
and other participants in the workshop "Future of Fission Theory",
York, UK (2019) for discussions motivating this work.

\end{document}